\documentclass[10pt, conference, letterpaper]{IEEEtran}
\ifCLASSINFOpdf
\else
\fi
\usepackage{graphicx}
\usepackage{amsmath}
\usepackage{amstext}
\usepackage{bm}
\usepackage{booktabs,cite,balance}
\usepackage{epstopdf}
\usepackage{multirow}

\newenvironment{sequation*}{\begin{equation*}\small}{\end{equation*}}

\long\def\symbolfootnote[#1]#2{\begingroup%
\def\thefootnote{\fnsymbol{footnote}}\footnote[#1]{#2}\endgroup}

\ifCLASSOPTIONcompsoc
\usepackage[nocompress]{cite}
\else
\usepackage{cite}
\fi

\begin{document}
\pagestyle{empty}

\bibliographystyle{IEEEtran}

\title{Information-Centric Wireless Networks with Mobile Edge Computing}
\author{Yuchen~Zhou\IEEEauthorrefmark{1}, F.~Richard~Yu\IEEEauthorrefmark{2}, Jian~Chen\IEEEauthorrefmark{1}, and Yonghong~Kuo\IEEEauthorrefmark{1}\\
\IEEEauthorblockA{\IEEEauthorrefmark{1}School of Telecomm. Eng., Xidian University, Xi'an 710071, P.R. China}
\IEEEauthorblockA{\IEEEauthorrefmark{2}Dept. of Systems and Computer Eng., Carleton University, Ottawa, ON, Canada}
\IEEEauthorblockA{\textit{Corresponding author: Jian Chen (jianchen@mail.xidian.edu.cn)}}}
\maketitle

\thispagestyle{empty}

\begin{abstract}
In order to better accommodate the dramatically increasing demand for data caching and computing services, storage and computation capabilities should be endowed to some of the intermediate nodes within the network. In this paper, we design a novel virtualized heterogeneous networks framework aiming at enabling content caching and computing. With the virtualization of the whole system, the communication, computing and caching resources can be shared among all users associated with different virtual service providers. We formulate the virtual resource allocation strategy as a joint optimization problem, where the gains of not only virtualization but also caching and computing are taken into consideration in the proposed architecture. In addition, a distributed algorithm based on alternating direction method of multipliers is adopted to solve the formulated problem, in order to reduce the computational complexity and signaling overhead. Finally, extensive simulations are presented to show the effectiveness of the proposed scheme under different system parameters.
\end{abstract}


\section{Introduction}
According to the investigation conducted in \cite{tassi2015resource}, the global video traffic will dominate the Internet usage in the future. Since heterogeneous devices request different kinds of video formats, resolutions, and bitrates, the existing video contents may need to be transformed to fit the network condition and the usage of different mobile devices. Therefore, transcoding technology is necessary for transforming the current video version into a suitable one, which can be played and matches the screen size of the devices. However, such a transcoding procedure is computation-intensive so that it can be hardly executed on the mobile devices with limited resources. Thereby, a novel computing platform is desirable.\\
\indent \emph{Mobile edge computing} (MEC) is recognized as a promising paradigm in next generation wireless networks, enabling the cloud-computing capabilities in close proximity to mobile devices \cite{kumar2016vehicular}. With the physical proximity, MEC realizes a low-latency connection to a large-scale resource-rich computing infrastructure by offloading the computation task to an adjacent computing sever/cluster instead of relying on a remote cloud \cite{chen2015efficient}. Therefore, MEC is envisioned to provide computation services for mobile devices at anytime and anywhere by endowing radio access networks (RANs) with powerful computing capabilities \cite{zhang2016energy}.\\
\indent Since the prodigious amount of videos and the wide variety of video versions will certainly result in a large-scale distribution of video contents calling for tremendous resources, it is essential to have storage resources at some of the intermediate nodes within the network \cite{jin2015optimal}. \emph{In-network caching} can help efficient distribution of contents in wireless networks \cite{FYH15}. Compared to the traditional network paradigms with a general lack of content distribution information, the cache-enable heterogeneous networks (HetNets) can reduce the backhaul cost of the popular contents, increase the delivery probability of contents to mobile users, and support a highly efficient and scalable content retrieval.\\
\indent In view of the benefits from MEC and in-network caching, a novel framework integrated with these promising techniques is necessary to be designed for efficiently delivering the massive video contents in HetNets. In this paper, \emph{wireless network virtualization} is considered as a candidate technique for simplifying network management \cite{LY15,LY15m}. Through virtualization, wireless network infrastructure can be decoupled from their provided services, and various users with differentiated services requirements can dynamically share the same infrastructure, thereby maximizing the system utilization \cite{wang2016information}.\\
\indent Thanks to MEC and in-network caching, the computing and caching functions can be achieved in close proximity to mobile devices. However, although some excellent works have been done on MEC and in-network caching, these two areas have been addressed separately. Thus, how to integrate these two techniques, and efficiently allocate the limited resources to jointly optimize the utilities of computing, caching, and communication, remain to be an urgent issue.\\
\indent In this paper, we investigate the virtualized HetNets with MEC and in-network caching. Specifically, we design a novel virtualized HetNets framework aiming at enabling content caching and computing, in which the resources of communication, computing, and caching can be shared among users from different virtual networks. In this framework, we formulate the virtual resource allocation strategy as a joint optimization problem, where the gains of not only virtualization but also caching and computing are taken into consideration in the proposed HetNets virtualization architecture. A distributed algorithm, based on alternating direction method of multipliers (ADMM) \cite{boyd2011distributed}\cite{leinonen2013distributed}, is presented to solve the formulated problem with a lower computational complexity and a reduced signaling overhead. Simulations results are presented to show the performance improvements of the proposed scheme.

\indent The rest of this paper is organized as follows. In Section II, we introduce the proposed framework and formulate the virtual resource allocation scheme as an optimization problem. In Section III, we address the problem via a distributed ADMM-based algorithm. Simulation results are discussed in Section IV. Finally, we conclude this study in Section V.
\section{Virtual Resources Allocation with Mobile Edge Computing and In-Network Caching}
\subsection{System Model}
\begin{figure}
\centering
\includegraphics[width=3.7in]{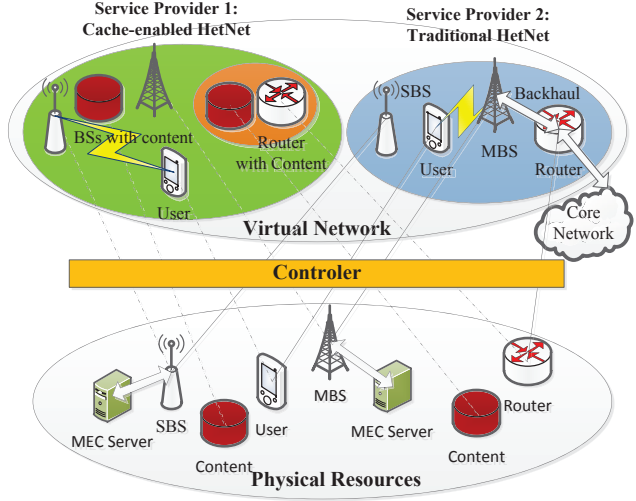}
\caption{Virtualized HetNets Framework.}
\label{fig1}
\end{figure}
\indent\textit{1): Virtual Heterogeneous Networks Model}\\
\indent As shown in Fig. 1, the virtual networks are generated according to the requests of service providers (SPs) by mobile virtual network operators (MVNO), since the quality-of-service (QoS) requirements may be different for each mobile user. In particular, some users, who want to compute the task such as face recognition, prefer to access to a virtual network without in-network caching function (i.e., SP2), because the contents of these kinds of computation tasks may be private and with very low reuse probability. In contrast, some computation tasks like video transcoding are better to be executed in the virtual network with in-network caching function (i.e., SP1). The virtual network with in-network caching function offers an opportunity to cache the popular contents before or after the execution of the computation tasks, which will significantly reduce the operation cost for delivering the video contents. It is assumed that the whole virtualization process is realized and controlled by a virtual wireless network controller \cite{ETSI2013NFV}. For simplicity, user mobility \cite{YL01} and handover \cite{MYL04,MYL07,YK07} are not considered in this paper. Each mobile user can connect to the virtual wireless networks logically, and subscribe to the required services from these virtual networks, while actually they connect to the physical networks.\\
\indent Considering that the HetNets are with multiple macro base stations (MBSs) and small base stations (SBSs) for serving multiple users, let $\mathcal{N}_m$ and $\mathcal{N}_s$ be the sets of MBSs and SBSs, and $\mathcal{N}=\mathcal{N}_m\cup\mathcal{N}_s=\{1,...,N\}$ and $\mathcal{U}=\{1,...,U\}$ be the sets of all BSs and users, respectively. It is assumed that each BS belongs to different infrastructure providers (InPs), and the licensed spectrum of each InP is orthogonal so that there is no interference among them. In addition, let $\mathcal{S}=\{1,...,S\}$ be the set of SPs. For each SP $s$, each assigned user is denoted by $u_s$, and $\mathcal{U}_s$ is the set of users belonging to SP $s$, where $\mathcal{U}=\cup_s\mathcal{U}_s$ and $\mathcal{U}_s\cap\mathcal{U}_{s'}=\phi$, $\forall {s'}\neq s$.\\
\indent In our business model, MVNO leases radio resource (e.g., spectrum) and backhaul bandwidth (e.g., data rate) from InPs, and slices them to virtual SPs. On the revenue side, MVNO charge the access fee of virtual network from users, which is defined as $\alpha_{u_s}$ per bps. The users, who have already paid the fee, can access to the virtual network for offloading their computation task. Besides, the fee for user $u_s$ to compute the task at BS $n$ is defined as $\phi_{u_s}$ per bps. Since the contents of the computation tasks may have the potential benefits to be cached, the backhaul cost, paid by MVNO and defined as $\gamma_n$ per bps, can be saved when users call for the contents which have already been cached at BS $n$. On the spending side, MVNO needs to dynamically pay for the usage of spectrum to InPs, which is defined as $\beta_n$ per Hz. Furthermore, MVNO also needs to pay the computation fee and caching fee to InPs, once there is a computation task need to be executed at the MEC server or the contents before and after the computation are valuable to be cached at BSs. In addition, with the increasingly rigid environmental standards and rising energy costs, there are great interests on the energy issues in wireless networks \cite{XYJL12,BYC12,YZX11,BYY15}. Therefore, we define the unit price of computation energy at BS $n$ as $\psi_n$ per J. The prices per unit of space to cache the contents before and after the computation at BS $n$ are denoted by $\Psi^n_{z_{u_s}}$ and $\Psi^n_{z_{u_s}'}$, where $z_{u_s}$ and $z_{u_s}'$ represent the contents before and after the computation.\\
\indent\textit{2): Computing Model}\\
\indent Assume each user has a computation task to be completed with a certain requirement of computation rate. Let $a_{u_s,n}$ denote the association indicator, where $a_{u_s,n}=1$ means that user $u_s$ associates with BS $n$ to compute the offloading task. Each user can associate to only one BS; thus
\begin{equation}
\sum\limits_{n\in\mathcal{N}}a_{u_s,n}=1, \forall s\in\mathcal{S},u_s\in\mathcal{U}_s.
\end{equation}
$b_{u_s,n}$ denotes the allocated bandwidth from BS $n$ to user $u_s$, and we have
\begin{equation}
\sum\limits_{s\in\mathcal{S}}\sum\limits_{u_s\in\mathcal{U}_s}a_{u_s,n}b_{u_s,n}\leq B_n, \forall n\in\mathcal{N},
\end{equation}
where $B_n$ is used to denote the spectrum bandwidth allocated to BS $n$. In order to ensure the data rate requirements of each user, we have
\begin{equation}
\sum\limits_{n\in\mathcal{N}}a_{u_s,n}b_{u_s,n}r_{u_s,n}\geq R^{\text{cm}}_{u_s}, \forall s\in\mathcal{S},u_s\in\mathcal{U}_s.
\end{equation}
where $R^{\text{cm}}_{u_s}$ is user $u_s$'s communication rate requirement in the corresponding QoS class. According to the Shannon bound, $r_{u_s,n}$, the achievable spectrum efficiency of user $u_s$ associating with BS $n$, can be easily obtained.\\
\indent Assume each computation task can be described in four terms as $T_{u_s}=\{z_{u_s},z'_{u_s},c_{u_s},R^{\text{cp}}_{u_s}\}, \forall s,u$. For the task $T_{u_s}$, $z_{u_s}$ and $z'_{u_s}$ respectively represent the sizes of the contents before and after the computation. $c_{u_s}$ denotes the computing ability required for accomplishing this task, which can be quantized by the amount of CPU cycles \cite{chen2015efficient}. $R^{\text{cp}}_{u_s}$ is the minimum computation rate required by user $u_s$.\\
\indent Let $e_{n}$ be the energy consumption for one CPU cycle at BS $n$. We denote $f_{u_s,n}$ as the computation capability of BS $n$ assigned to user $u_s$, which is quantized by the total number of CPU cycles per second \cite{chen2015efficient}. Then the computation execution time of the task at BS $n$ can be easily obtained as $t_{u_s,n}=\frac{c_{u_s}}{f_{u_s,n}}$. Therefore, the computation rate (i.e., the amount of bits computed during one second) of BS $n$ to compute task $T_{u_s}$ can be equivalent to $R_{u_s,n}=\frac{z_{u_s}}{t_{u_s,n}}=\frac{f_{u_s,n}z_{u_s}}{c_{u_s}}$, and the total energy consumption used for computing task $T_{u_s}$ at BE $n$ can be calculated as $E_{u_s,n}=c_{u_s}e_{n}$.\\
\indent Since each user has the requirement for computation rate,
\begin{equation}
\sum\limits_{n\in\mathcal{N}}a_{u_s,n}R_{u_s,n}\geq R^{\text{cp}}_{u_s}, \forall s\in\mathcal{S},u_s\in\mathcal{U}_s.
\end{equation}
Moreover, it should be noted that the computation ability at each BS is limited; thus
\begin{equation}
\sum\limits_{s\in\mathcal{S}}\sum\limits_{u_s\in\mathcal{U}_s}a_{u_s,n}\leq D_{n}, \forall n\in\mathcal{N},
\end{equation}
where $D_n$ is the maximum amount of tasks simultaneously executed on the MEC server of BS $n$.\\
\indent\textit{3): Caching Model}\\
\indent For each BS, they can determine whether to cache the content sent by users before or after the computation, according to the popularity distribution of each content. The caching strategy can be controlled by two binary parameter $x^1_{u_s,n}$ and $x^2_{u_s,n}$. If BS $n$ caches the original content, $x^1_{u_s,n}=1$; otherwise $x^1_{u_s,n}=0$. If BS $n$ caches the computed content, $x^2_{u_s,n}=1$; otherwise $x^2_{u_s,n}=0$. It
should be noted that the storage of BS $n$ may be limited. Thus, the cached content cannot be larger than the remaining space $Z_n$ of BS $n$, which can be expressed as
\begin{equation}
\sum\limits_{s\in\mathcal{S}}\sum\limits_{u_s\in\mathcal{U}_s}a_{u_s,n}(x^1_{u_s,n}z_{u_s}+x^2_{u_s,n}z_{u_s}')\leq Z_{n}, \forall n\in\mathcal{N}.
\end{equation}
\indent In this paper, it is assumed that the popularity distribution is represented by a vector $\bm p=[p_1,p_2,...,p_F]$, where $F$ types of contents with diverse popularity are distributed in the networks. That is, each content $f$ is requested by each mobile user independently with the probability $p_f$. Generally, $\bm p$ is modeled as the Zipf distribution \cite{li2016pricing}, which can be expressed as
\begin{equation}
p_f=\frac{1/{f^\epsilon}}{\sum\limits_{f=1}^{F}1/{f^\epsilon}}, \forall f,
\end{equation}
where the exponent $\epsilon$ is a positive value and can characterizes the content popularity. For our business model, $p_{z_{u_s}}$ and $p_{z_{u_s}'}$ can be directly derived from $p_f$ if the content sent by user $u_s$ is known. Afterwards, the gains of the expected saved backhaul bandwidth through caching contents $z_{u_s}$ and $z_{u_s}'$ can be respectively calculated as $g_{z_{u_s}}=\frac{p_{z_{u_s}}z_{u_s}}{T_{z_{u_s}}}$ and $g_{z_{u_s}'}=\frac{p_{z_{u_s}'}z_{u_s}'}{T_{z_{u_s}'}}$, where $T_{z_{u_s}}$ and $T_{z_{u_s}'}$ are the time durations for downloading the required contents through backhaul.
\subsection{Problem Formulation}
In this subsection, an optimization problem is formulated to maximize the aggregate utility of the MVNO system. The optimization problem is mathematically modeled as
\begin{eqnarray*}
&&OP1:\max\limits_{\mbox{\tiny$\begin{array}{c}\{a_{u_s,n},b_{u_s,n},\\ x^1_{u_s,n},x^2_{u_s,n}\}\end{array}$}} \sum\limits_{s\in\mathcal{S}}\sum\limits_{u_s\in\mathcal{U}_s}\sum\limits_{n\in\mathcal{N}}U_{u_s,n}\\
&&s.t.:(1)(2)(3)(4)(5)(6)\\
\end{eqnarray*}
where $U_{u_s,n}$ is the potential utility of user $u_s$ associating with BS $n$, and it can be defined as
\begin{equation}
\begin{array}{r}
U_{u_s,n}=a_{u_s,n}(\alpha_{u_s}b_{u_s,n}r_{u_s,n}-\beta_{n}b_{u_s,n})\\
+a_{u_s,n}(\phi_{u_s}R_{u_s,n}-\psi_n E_{u_s,n})\\
+a_{u_s,n}x^1_{u_s,n}(\gamma_ng_{z_{u_s}}-\Psi^n_{z_{u_s}}z_{u_s})\\
+a_{u_s,n}x^2_{u_s,n}(\gamma_ng_{z_{u_s}'}-\Psi^n_{z_{u_s}'}z_{u_s}').
\end{array}
\end{equation}
Here, $\alpha_{u_s}b_{u_s,n}r_{u_s,n}$ denotes the gain of user data rate, $\beta_{n}b_{u_s,n}$ is the cost of consumed radio bandwidth, $\phi_{u_s}R_{u_s,n}$ denotes the gain of computation rate, $\psi_n E_{u_s,n}$ is the cost of consumed computation energy, $\gamma_ng_{z_{u_s}}$ and $\gamma_ng_{z_{u_s}'}$ are the gains achieved on the saved backhaul bandwidth from caching the contents $z_{u_s}$ and $z_{u_s}'$, and $\Psi^n_{z_{u_s}}z_{u_s}$ and $\Psi^n_{z_{u_s}'}z_{u_s}'$ are the costs of caching the contents $z_{u_s}$ and $z_{u_s}'$, respectively.
\subsection{Problem Reformulation}
It is obvious that the formulated mixed discrete and non-convex optimization problem is a NP-hard problem \cite{fooladivanda2013joint}. A relaxation of the binary conditions of $a_{u_s,n}$, $x^1_{u_s,n}$, and $x^2_{u_s,n}$ constitutes the first step to solve the problem OP1, where $a_{u_s,n}$, $x^1_{u_s,n}$, and $x^2_{u_s,n}$ are relaxes to be real value variables as $0\leq a_{u_s,n}\leq 1$, $0\leq x^1_{u_s,n}\leq 1$ and $0\leq x^2_{u_s,n}\leq 1$. The relaxed $a_{u_s,n}$ is sensible and meaningful to be a time sharing factor representing the ratio of time for user $u_s$ to associate with BS $n$ in order to offload and compute the offloading task. The relaxed $x^1_{u_s,n}$ and $x^2_{u_s,n}$ can be also interpreted as the time fractions for sharing one unit cache of BS $n$.\\
\indent However, even after the relaxation of the variables, the problem is still non-convex due to the multiplication of the variables. Thus, a second step is necessary for further simplifying the problem to make it tractable and solvable.\\
\indent\textit{Proposition 4.1:} If we define $\widetilde x^1_{u_s,n}=a_{u_s,n}x^1_{u_s,n}$, $\widetilde x^2_{u_s,n}=a_{u_s,n}x^2_{u_s,n}$, and $\widetilde b_{u_s,n}=a_{u_s,n}b_{u_s,n}$, there exists an equivalent formulation of problem OP1 as follows:
\begin{eqnarray*}
&&OP2:\max\limits_{\mbox{\tiny$\begin{array}{c}\{a_{u_s,n},\widetilde b_{u_s,n},\\ \widetilde x^1_{u_s,n},\widetilde x^2_{u_s,n}\}\end{array}$}} \sum\limits_{s\in\mathcal{S}}\sum\limits_{u_s\in\mathcal{U}_s}\sum\limits_{n\in\mathcal{N}}\widetilde U_{u_s,n}\\
&&s.t.:C1:\sum\limits_{n\in\mathcal{N}}a_{u_s,n}=1, \forall s\in\mathcal{S},u_s\in\mathcal{U}_s\\
&&C2:\sum\limits_{s\in\mathcal{S}}\sum\limits_{u_s\in\mathcal{U}_s}\widetilde b_{u_s,n}\leq B_n, \forall n\in\mathcal{N}\\
&&C3:\sum\limits_{n\in\mathcal{N}}\widetilde b_{u_s,n}r_{u_s,n}\geq R^{\text{cm}}_{u_s}, \forall s\in\mathcal{S},u_s\in\mathcal{U}_s\\
&&C4:\sum\limits_{n\in\mathcal{N}}a_{u_s,n}R_{u_s,n}\geq R^{\text{cp}}_{u_s}, \forall s\in\mathcal{S},u_s\in\mathcal{U}_s\\
&&C5:\sum\limits_{s\in\mathcal{S}}\sum\limits_{u_s\in\mathcal{U}_s}a_{u_s,n}\leq D_{n}, \forall n\in\mathcal{N}\\
&&C6:\sum\limits_{s\in\mathcal{S}}\sum\limits_{u_s\in\mathcal{U}_s}(\widetilde  x^1_{u_s,n}z_{u_s}+ \widetilde x^2_{u_s,n}z_{u_s}')\leq Z_{n}, \forall n\in\mathcal{N}\\
\end{eqnarray*}
\indent The relaxed problem OP1 can be directly recovered through substituting the variables $\widetilde x^1_{u_s,n}=a_{u_s,n}x^1_{u_s,n}$, $\widetilde x^2_{u_s,n}=a_{u_s,n}x^2_{u_s,n}$, and $\widetilde b_{u_s,n}=a_{u_s,n}b_{u_s,n}$ into problem OP2. If $a_{u_s,n}=0$, $b_{u_s,n}=0$ certainly holds due to the optimality. Obviously, there is no need for BS $n$ to allocate any resource to a user when the user does not associate with BS $n$.\\
\indent Now problem OP2 is transformed as a convex problem. However, the signaling overhead will be prohibitively large if a centralized algorithm is used to solve the problem, because finding out the optimal solution requires all the channel state information (CSI) and content distribution information. Therefore, a distributed optimization algorithm executed on each BS is necessary to be designed for practical implementing. However, because of the constraints $C1, C3$, and $C4$, problem OP2 is not separable to be executed on each BS. Thus, the coupling has to be decoupled appropriately, which will be discussed in Section III. To lighten the notation, from now on, $u$ is used to denote each user instead of $u_s$.
\section{Resource Allocation via Alternating Direction Method of Multipliers}
In order to decouple the coupling variables, the local copies of $\{a_{u,n}\}$ and $\{\widetilde b_{u,n}\}$ at BS $n$ is introduced as $\{\widehat a^n_{u,k}\}$ and $\{\widehat b^n_{u,k}\}$, respectively. With the local vectors $\{\widehat a^n_{u,k}\}$ and $\{\widehat b^n_{u,k}\}$, a feasible local variable set for each BS $n$ can be defined as
\begin{equation}
\label{feasible set}
\mathcal{X}_n=\left\{ {\begin{array}{*{20}{c}}
\{\widehat a^{n}_{u,k}\}\\
\{\widehat b^n_{u,k}\}
\end{array}\left| \begin{array}{l}
\sum\limits_{k\in\mathcal{N}}\widehat a^n_{u,k}=1, \forall u\\
\sum\limits_{u\in\mathcal{U}}\widehat b^n_{u,k}\leq B_k, \forall k\\
\sum\limits_{k\in\mathcal{N}}\widehat b^n_{u,k}r_{u,k}\geq R^{\text{cm}}_{u}, \forall u\\
\sum\limits_{k\in\mathcal{N}}\widehat a^n_{u,k}R_{u,k}\geq R^{\text{cp}}_{u}, \forall u\\
\sum\limits_{u\in\mathcal{U}}\widehat a^n_{u,k}\leq D_{k}, \forall k\\
\sum\limits_{u\in\mathcal{U}}(\widetilde x^{1}_{u,k}z_{u}+ \widetilde x^{2}_{u,k}z_{u}')\leq Z_{k}, \forall k\\
\end{array} \right.} \right\},
\end{equation}
and an associated local utility function can be expressed as
\begin{equation}
\mathcal{y}_n=\left\{ \begin{aligned} &-\sum\limits_{u\in\mathcal{U}}\widehat U_{u,n}, (\{\widehat a^{n}_{u,k}\}, \{\widetilde x^{1}_{u,k}\}, \{\widetilde x^2_{u,k}\}, \{\widehat b^n_{u,k}\})\in\mathcal{X}_n \\ &0,\quad\quad\text{Otherwise} \end{aligned} \right.
\end{equation}
With this notation, the global consensus problem of the problem OP2 can be shown as follows:
\begin{eqnarray*}
&&OP3:\min \mathcal{Y}(\{\widehat a^{n}_{u,k}\}, \{\widetilde x^{1}_{u,k}\}, \{\widetilde x^2_{u,k}\}, \{\widehat b^n_{u,k}\})=\\
&&\quad\quad\quad\sum\limits_{n\in\mathcal{N}}\mathcal{y}_n(\{\widehat a^{n}_{u,k}\}, \{\widetilde x^{1}_{u,k}\}, \{\widetilde x^2_{u,k}\}, \{\widehat b^n_{u,k}\})\\
&&s.t.: \{\widehat a^{n}_{u,k}\}=\{a_{u,k}\}, \{\widehat b^{n}_{u,k}\}=\{\widetilde b_{u,k}\}, \forall n,u,k
\end{eqnarray*}\\
\indent Obviously, now the objective function is separable across each BS. The initial step of ADMM to solve the problem OP3 is the formulation of an augmented Lagrangian $\mathcal{L}_{\rho}(\{\widehat {\bm a},\widetilde {\bm x}^1,\widetilde {\bm x}^2,\widehat {\bm b}\},\{\bm a,\widetilde {\bm b}\},\{\bm\mu,\bm\nu\})$ with corresponding global consensus constrains. Here, $\widehat {\bm a}=\{\widehat a^{n}_{u,k}\}$, $\widetilde {\bm x}^1=\{\widetilde x^{1}_{u,k}\}$, $\widetilde {\bm x}^2=\{\widetilde x^{2}_{u,n}\}$, $\widehat {\bm b}=\{\widehat b^n_{u,k}\}$, $\bm a=\{a_{u,k}\}$, and $\widetilde {\bm b}=\{\widetilde b_{u,k}\}$. The augmented Lagrangian can be derived as \cite{boyd2011distributed}
\begin{equation}
\begin{array}{r}
\mathcal{L}_{\rho}(\{\widehat {\bm a},\widetilde {\bm x}^1,\widetilde {\bm x}^2,\widehat {\bm b}\},\{\bm a,\widetilde {\bm b}\},\{\bm\mu,\bm\nu\})=\sum\limits_{n\in\mathcal{N}}\mathcal{y}_n(\widehat {\bm a}^{n}, \widetilde {\bm x}^1, \widetilde {\bm x}^2, \widehat {\bm b}^n)+\\
\sum\limits_{n\in\mathcal{N}}\sum\limits_{\substack{u\in\mathcal{U}\\k\in\mathcal{N}}}\mu^n_{u,k}(\widehat a^n_{u,k}- a_{u,k})+\frac{\rho}{2}\sum\limits_{n\in\mathcal{N}}\sum\limits_{\substack{u\in\mathcal{U}\\k\in\mathcal{N}}}(\widehat a^n_{u,k}- a_{u,k})^2+\\
\sum\limits_{n\in\mathcal{N}}\sum\limits_{\substack{u\in\mathcal{U}\\k\in\mathcal{N}}}\nu^n_{u,k}(\widehat b^n_{u,k}-\widetilde  b_{u,k})+\frac{\rho}{2}\sum\limits_{n\in\mathcal{N}}\sum\limits_{\substack{u\in\mathcal{U}\\k\in\mathcal{N}}}(\widehat b^n_{u,k}-\widetilde b_{u,k})^2,
\end{array}
\end{equation}
where $\rho$ is the penalty parameter, and $\bm\mu=\{\mu^n_{u,k}\}$ and $\bm\nu=\{\nu^n_{u,k}\}$ are the dual variables.\\
\indent According to the iteration of AMDD with consensus constraints, the process for solving the problem OP3 consists of the following several steps:\\
\indent\textit{Step 1: $\{\widehat {\bm a}^{n},\widetilde {\bm x}^{1},\widetilde {\bm x}^{2},\widehat {\bm b}^{n}\}-update$:} In this step, the problem OP3 can be completely decoupled into $N$ specific subproblems, and each of the subproblems can be solved locally and separately at BSs. BS $n$ solves the following optimization problem at iteration $[i]$:
\begin{equation}
\begin{array}{r}
\{\widehat {\bm a}^n,\widetilde {\bm x}^{1},\widetilde {\bm x}^{2},\widehat {\bm b}^{n}\}^{[i+1]}_{n\in\mathcal{N}}:=\arg\min\{\mathcal{y}_n(\widehat {\bm a}^{n}, \widetilde {\bm x}^{1}, \widetilde {\bm x}^{2}, \widehat {\bm b}^n)\\
+\sum\limits_{\substack{u\in\mathcal{U}\\k\in\mathcal{N}}}\mu^{n[i]}_{u,k}(\widehat a^n_{u,k}- a^{[i]}_{u,k})+\frac{\rho}{2}\sum\limits_{\substack{u\in\mathcal{U}\\k\in\mathcal{N}}}(\widehat a^n_{u,k}-a^{[i]}_{u,k})^2\\
+\sum\limits_{\substack{u\in\mathcal{U}\\k\in\mathcal{N}}}\nu^{n[i]}_{u,k}(\widehat b^n_{u,k}-\widetilde b^{[i]}_{u,k})+\frac{\rho}{2}\sum\limits_{\substack{u\in\mathcal{U}\\k\in\mathcal{N}}}(\widehat b^n_{u,k}-\widetilde b^{[i]}_{u,k})^2\}.
\end{array}
\end{equation}
In this paper, the primal dual interior-point method, which is able to provide an efficient way for solving convex problems \cite{boyd2004convex}, is used to find out the optimal solution of the problem. Due to the limited space, the details of the procedure are omitted here.\\
\indent\textit{Step 2: $\{{\bm a}, \widetilde {\bm b}\}-update$:} In the second step, $\bm a$ and $\widetilde {\bm b}$ can be updated according to
\begin{equation}
\begin{array}{r}
{\bm a}^{[i+1]}:=\arg\min\sum\limits_{n\in\mathcal{N}}\sum\limits_{\substack{u\in\mathcal{U}\\k\in\mathcal{N}}}\mu^{n[i]}_{u,k}(\widehat a^{n[i+1]}_{u,k}-a_{u,k})\\
+\frac{\rho}{2}\sum\limits_{n\in\mathcal{N}}\sum\limits_{\substack{u\in\mathcal{U}\\k\in\mathcal{N}}}(\widehat a^{n[i+1]}_{u,k}-a_{u,k})^2,\\
\widetilde {\bm b}^{[i+1]}:=\arg\min\sum\limits_{n\in\mathcal{N}}\sum\limits_{\substack{u\in\mathcal{U}\\k\in\mathcal{N}}}\nu^{n[i]}_{u,k}(\widehat b^{n[i+1]}_{u,k}-\widetilde b_{u,k})\\
+\frac{\rho}{2}\sum\limits_{n\in\mathcal{N}}\sum\limits_{\substack{u\in\mathcal{U}\\k\in\mathcal{N}}}(\widehat b^{n[i+1]}_{u,k}-\widetilde b_{u,k})^2.
\end{array}
\end{equation}
Because we have added the quadratic regularization term to the augmented Lagrangian (11), the unconstrained problems (13) are strictly convex with respect to ${\bm a}$ and $\widetilde {\bm b}$.\\
\indent\textit{Step 3: $\{\bm\mu,\bm\nu\}-update$:} This step shows the updating of dual variables, which can be represented as
\begin{equation}
\begin{array}{r}
\bm\mu^{n[i+1]}:=\bm\mu^{n[i]}+\rho(\widehat {\bm a}^{n[i+1]}-{\bm a}^{[i+1]}),\\
\bm\nu^{n[i+1]}:=\bm\nu^{n[i]}+\rho(\widehat {\bm b}^{n[i+1]}-\widetilde {\bm b}^{[i+1]}).
\end{array}
\end{equation}
Here, the augmented Lagrangian parameter $\rho$ is used as a step size to update the dual variables.\\
\indent\textit{Step 4: Algorithm Stopping Criterion:} According to \cite{boyd2011distributed}, the residual for the primal feasibility condition of BS $n$ at iteration $[i]$ should be small enough so that
\begin{equation}
\begin{array}{r}
||\widehat {\bm a}^{n[i+1]}-{\bm a}^{[i+1]}||_2\leq\upsilon_{pri},\quad
||\widehat {\bm b}^{n[i+1]}-\widetilde {\bm b}^{[i+1]}||_2\leq\upsilon_{pri}.
\end{array}
\end{equation}
Moreover, the residual for the first dual feasibility condition at iteration $[i+1]$ should be small enough so that
\begin{equation}
\begin{array}{r}
||{\bm a}^{[i+1]}-{\bm a}^{[i]}||_2\leq\upsilon_{dual},\quad
||\widetilde {\bm b}^{[i+1]}-\widetilde {\bm b}^{[i]}||_2\leq\upsilon_{dual}.
\end{array}
\end{equation}
Here, $\upsilon_{pri}>0$ and $\upsilon_{dual}>0$, called as the feasibility tolerances of the primal feasibility and dual feasibility conditions, respectively. Finally, after obtaining the optimum solution, the binary recovery can be viewed as computing the marginal benefit for each user $u$ \cite{LYJ16}.
\section{Simulation Results and Discussions}
\begin{table}
\renewcommand{\arraystretch}{1.3}
\caption{Parameter Values}
\label{tab:1}
\centering
\begin{tabular}{c|c||c|c}
\hline
Parameter & Value & Parameter & Value\\
\hline
$R_u^{cm}$ & $10^5$bps & $R_u^{cp}$ & $10^5$bps \\
$e_n$ & $1$W/GHz & $D_n$ & $10$\\
$\beta_n$ & [1,3] units/KHz & $\psi_n$ & [40,80]*$10^{-6}$ units/J\\
$\Psi^n_{z_{u_s}}$ & [10,20] units/Mb & $\Psi^n_{z'_{u_s}}$ & [10,20] units/Mb\\
$\alpha_n$ & $10$units/Mbps & $\phi_{u_s}$ & $100$units/bps\\
$\gamma_1$ & $10$units/Mbps & $\gamma_2$ & $12$units/Mbps\\
noise & -174dBm & power & 27dBm\\
\hline
\end{tabular}
\end{table}
We assume that SBSs and users are randomly distributed within the covered area of the MBS, and all the channel coefficients are distributed as $\mathcal {CN}(0,\frac{1}{(1+d)^{\alpha}})$ with a path loss exponent $\alpha=4$, where $d$ is the distance between each mobile user and BS. In addition, there are two SPs, two BSs, one MVNO, and the bandwidth of each BS is normalized. The values of the rest of parameters are summarized in Table I.\\
\begin{figure}
\centering
\includegraphics[width=3.7in]{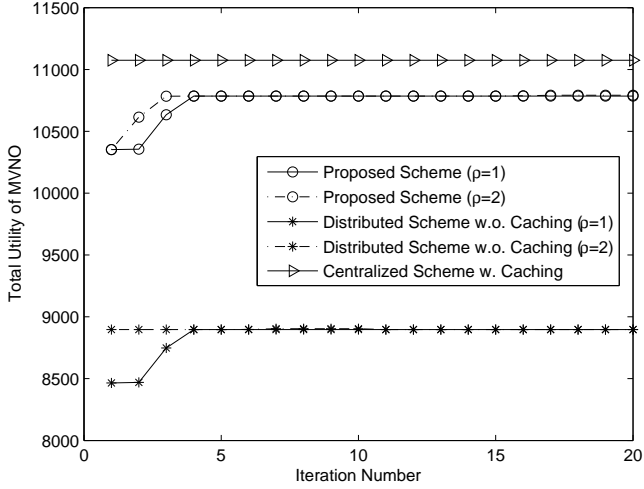}
\caption{Convergence of the algorithms. (The total number of users is 8. The sizes of each content are randomly distributed within 1Mb to 4Mb. The computing ability is distributed within 100Megacycles to 1300Megacycles. The computation capability of two BSs are 10GHz and 5GHz, and the cache spaces of two BSs are 10Mb and 5Mb.)}
\label{fig2}
\end{figure}
\indent Fig. 2 shows the convergence of the proposed scheme under different values of $\rho$. All schemes are able to converge to a stable solution rapidly, and the proposed scheme with different values of $\rho$ can eventually converge to a same value of the total utility. However, a higher value of $\rho$ will result in a higher rate of convergence. Thus, in the following simulations, we set $\rho=2$. Furthermore, we can observe that the proposed scheme performs better than the distributed scheme without caching function. Although there is a performance gap from the centralized scheme, the advantage of the proposed scheme is the reduced signal overhead for the exchange of the content distribution information and the CSI.\\
\begin{figure}
\centering
\includegraphics[width=3.7in]{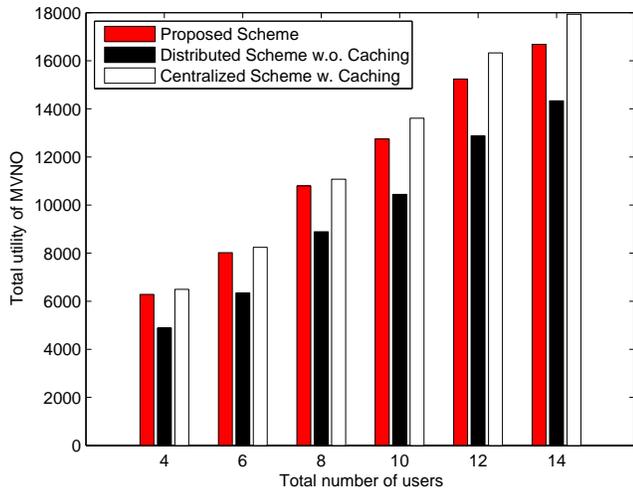}
\caption{Total utility of MVNO with different numbers of users. (The sizes of each content are randomly distributed within 1Mb to 4Mb. The computing ability is distributed within 100Megacycles to 1300Megacycles. The computation capability of two BSs are 10GHz and 5GHz, and the cache spaces of two BSs are 10Mb and 5Mb.)}
\label{fig3}
\end{figure}
\indent Fig. 3 illustrates the total utility of different schemes with respect to the different values of total number of users. As the number of users increases, the total utilities of all the schemes continue to grow. The main reason for the performance of the distributed scheme without caching function being worse than the proposed scheme, is that the popular contents cannot be cached at BSs so that there is no caching revenue when some users call for the previous contents. On the other hand, in the proposed scheme, if the computed contents required by users have already been cached at the associated BSs, the BSs does not need to compute the offloading contents, which will certainly contribute to increasing the computation revenue.\\
\balance
\section{Conclusion and Future Work}
In this paper, we studied virtual resource allocation for communication, computing, and caching in the designed virtualized HetNets framework. The allocation strategy was formulated as a joint optimization problem, considering the gains of not only virtualization but also caching and computing. In addition, a distributed ADMM-based algorithm was introduced to decouple the coupling variables and then split the optimization problem into several subproblems. Simulation results were presented to show the convergence and performance of the proposed scheme. Future work is in progress to consider software-defined networking (SDN) in the proposed framework.\\
\section*{Acknowledgment}
This work is jointly supported by the National Natural Foundation of China (Grant No. 61601347) and the `111' project of China (Grant No. B38038).

\bibliographystyle{IEEEtran}
\bibliography{reference}

\end{document}